\documentclass[aps,prd,twocolumn,superscriptaddress,floatfix]{revtex4}
\usepackage{amsmath, amsfonts}
\usepackage{graphicx}
\newcommand{\beq}{\begin{equation}}
\newcommand{\eeq}{\end{equation}}
\newcommand{\beqn}{\begin{eqnarray}}
\newcommand{\eeqn}{\end{eqnarray}}

\begin{document}
\title{Filling the disk hollow following binary black hole merger: \\ 
The transient accretion afterglow}
\date{\today}
\author{Stuart~L.~Shapiro}
\altaffiliation{Also Department of Astronomy and NCSA, University of
  Illinois at Urbana-Champaign, Urbana, IL 61801}
\affiliation{Department of Physics, University of Illinois at
  Urbana-Champaign, Urbana, IL 61801}

\begin{abstract}
Tidal torques from a binary black hole (BHBH) empty out the central regions in 
any circumbinary gaseous accretion disk. The balance between tidal
torques and viscosity maintain the inner edge of
the disk at a radius $r \sim 1.5a - 2a$, where $a$ is the
binary semimajor axis. Eventually, the inspiraling binary 
decouples from disk and merges, leaving behind a 
central hollow (``donut hole'') in the disk orbiting the remnant black hole.  
We present a simple, time-dependent, Newtonian calculation that follows the
secular (viscous) evolution of the disk as it fills up the hollow  
down to the black hole innermost stable circular orbit and then relaxes to 
stationary equilibrium. We use our model to 
calculate the electromagnetic radiation (``afterglow'') 
spectrum emitted during this transient accretion epoch. 
Observing the temporal increase in the total 
electromagnetic flux and the hardening of the spectrum as the donut hole fills 
may help confirm a BHBH merger detected by a gravitational wave interferometer. 
We show how the very existence of the initial hollow can lead to  
super-Eddington accretion during this secular phase if 
the rate is not very far below Eddington prior to decoupling.  Our model,
though highly idealized, may be useful in establishing some of 
the key parameters, thermal emission features and scalings that characterize 
this transient. It can serve as a guide in the design 
and calibration of future radiation-magnetohydrodynamic simulations in 
general relativity.
\end{abstract}
\maketitle

\section{Introduction}

Binary black holes (BHBHs) are among the most promising sources 
of gravitational waves detectable by gravitation-wave detectors 
such as LIGO~\cite{LIGO1,LIGO2}, VIRGO~\cite{VIRGO1,VIRGO2}, 
GEO~\cite{GEO}, and TAMA~\cite{TAMA1,TAMA2}, as
well as by the proposed space-based interferometers LISA~\cite{LISA}, 
BBO~\cite{bbo-nasa} and DECIGO~\cite{DECIGO}. The development of
stable algorithms to integrate Einstein's field equations of general
relativity numerically in $3+1$ dimensions, 
such as the BSSN formalism~\cite{ShiN95,BauS99}
and the generalized harmonic approach~\cite{Pre05a}, together with the 
identification of suitable gauge conditions, enabled
several pioneering simulations that demonstrated how to track the late 
inspiral, plunge and and merger of a binary black hole (BHBH) in 
vacuum~\cite{Pre05b,CamLPZ06,BakCCKV06}. 
More refined follow-up simulations of these strong-field, 
late phases, joined onto to analytic, post-Newtonian calculations of the 
early inspiral epoch~\cite{BlaFIS08}, are now cable of 
producing accurate gravitational 
waveforms for merging BHBH binaries with companions spanning a range of 
mass ratios and spins. These theoretical waveforms will be important as 
templates both for the detection of BHBH binaries and for the determination 
of their physical parameters, such as the masses and spins of the 
binary companions and the black hole remnant. 

With the binary BHBH merger problem in vacuum well in hand, it is now
important to turn to the problem of BHBH coalescence in astrophysically 
realistic gaseous enviroments in general relativity. 
When the orbit of the BHBH binary is 
sufficiently close, the ambient gas should have little
influence on the BHBH dynamics, but the impact of the black hole on the gas 
will be considerable. In particular, the capture, shock heating and accretion
of gas may result in appreciable electromagnetic radiation. Such 
an ``afterglow'' could provide another observable signature of BHBH 
coalescence in addition to gravitational waves. Electromagnetic radiation 
can also serve as a useful probe of the gas in galaxy cores or
in other environments where mergers may take place, as well as a probe 
of the physics of black hole accretion. 

\begin{figure*}
\includegraphics[width=17cm]{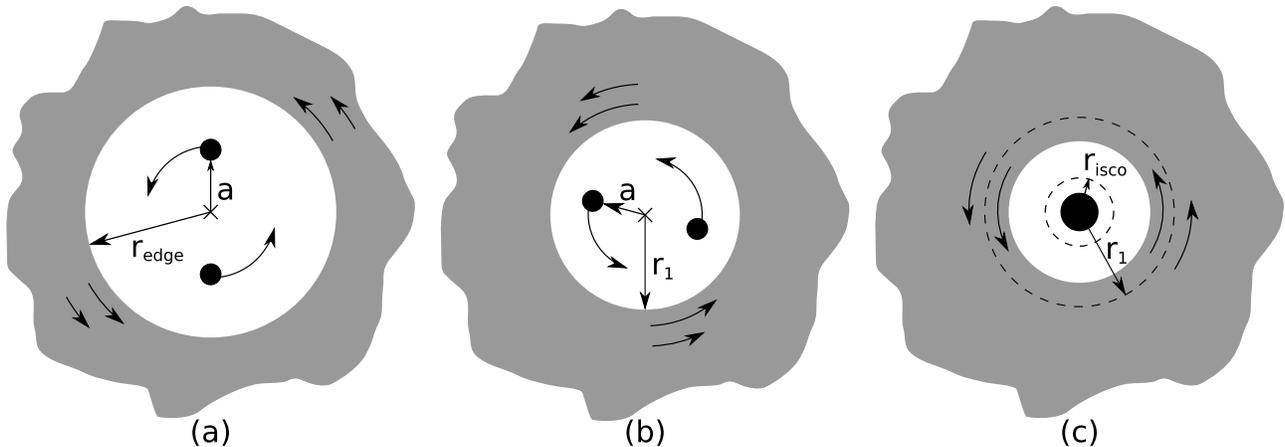}
\caption{Schematic figure of a Keplerian accretion disk orbiting a coalescing 
binary black hole. {\it (a)} The balance between tidal torques and viscosity 
maintains the inner edge of the disk at $r_{\rm edge} \sim 1.5a - 2a$, where
$a$ is the binary semimajor axis.  {\it (b)} Eventually the binary reaches a
critical separation at which its inspiral timescale due to gravitational
wave emission equals the viscous timescale in the disk. {\it (c)} The disk then
decouples from the binary, which coalesces, leaving a central black hole 
remnant.  Viscosity in the disk drives gas toward the remnant, filling up 
the hollow between the decoupling radius $r_1$ and the ISCO 
radius $r_{\rm isco}$.}
\label{fig:hollow}
\end{figure*}

The effect of a BHBH system on a rotationally supported, geometrically
thin, circumbinary disk has already triggered several important investigations,
most in Newtonian gravitation. One key finding is that the Bardeen \& 
Petterson mechanism~\cite{BarP75} will align the disk to  
the orbital plane of the binary on a viscous timescale~\cite{LarP97,IvaPP99}.
Another result, first appreciated in the context of disks around 
pre-main-sequence binaries and protoplanets orbiting a central star, 
is that the disk is truncated at an inner edge $r_{\rm edge}$ where 
gravitational tidal torques arising from the binary and 
viscous stresses in the disk balance each other~\cite{ArtL94}. 
As a result, a hollowed-out region (``donut hole") forms in the 
disk, causing gas accretion onto the black holes to remain
well below the rate it would have in the absence of tidal torques. 
During the binary inspiral, viscosity in the disk keeps the ratio of the 
inner edge radius to the semimajor axis $a$ of the binary roughly constant: 
$r_{\rm edge}/a \sim 1.5 - 2$~\cite{ArtL94,GunK02,EscLCM05,MacM08}.
Eventually, the binary separation shrinks sufficiently that
the binary inspiral timescale due to gravitational wave emission becomes 
shorter than the viscous timescale in the disk. At this point, the 
binary ``decouples'' from the disk, and the inspiral proceeds all 
the way to plunge and merger while the disk barely 
evolves~\cite{ArmN02,LiuWC03,MilP05}. The black hole remnant quickly 
settles down to a quasistationary state, after which the disk gas begins 
to drift inward on a secular, viscous timescale, filling the hollow down to the
innermost stable circular orbit (ISCO) of the remnant black hole at 
$r_{\rm isco}$.  
Eventually the disk settles down into quasistationary 
equilibrium and accretes at a steady rate onto the remnant. The 
scenario described here is summarized schematically in 
Fig.~\ref{fig:hollow}.

In this paper we follow the {\it secular} 
viscous evolution of a gaseous accretion disk following 
decoupling.  We perform a simple, Newtonian calculation to determine the 
time-dependent inspiral and structure of a geometrically thin disk 
as it fills up the hollow region and flows into the black hole remnant. 
We also calculate the associated thermal radiation spectrum of the disk 
from the time of decoupling to the 
the establishment of disk equilibrium. Our model is highly idealized in that
we assume a radiation spectrum at each radius given by thermal 
black body emission at the local effective temperature. In a
numerical example chosen to illustrate the model 
we also adopt a constant (turbulent) viscosity in space and time.
While idealized and largely pedagogical, the model 
provides a reasonable first approximation 
to the relevant time and length scales characterizing such a 
scenario, as well as the typical magnitudes and frequencies of the 
associated accretion-driven electromagnetic afterglow. The model may 
be useful in designing and calibrating future, more detailed, 
simulations that employ more realistic microphysics and relativistic 
gravitation.  Finally, adopting such a simple description for this transient
yields a calculation that exhibits considerable scale freedom. 
In particular, the time-dependent solution to the nondimensional 
evolution equation that governs the gas inflow
is uniquely determined by specifying only two free parameters in the 
initial data, the radius of the disk at decoupling, $r_1$ and the 
BH remnant ISCO, $r_{\rm isco}$. 
All other parameters may be scaled out of the problem. Moreover, 
given these two parameters, the equation does not depend on any further 
details of the disk microphysics.

Several recent calculations and hydrodynamic simulations have studied the 
{\it prompt} response of the circumbinary gas disk to gravitational wave-driven 
BHBH mass loss and remnant recoil following merger
(see, e.g., ~\cite{OneMBRS09,SchK08,CorHM09,RosLAPK09,AndLMN09} and 
references therein). 
The simulations treat effects that can occur on a hydrodynamical 
timescale in the absence of kinematic (shear) viscosity. They arise 
from those perturbations of the disk that are induced by the time-varying 
gravitational field of the binary just prior to 
merger and the recoiling remnant after merger. 
In fact, viscosity is typically ignored in these simulations. 
By contrast, the scenario considered below 
focusses on the slower, secular release of energy that ultimately accompanies 
the viscosity-driven inflow and accretion of the bound, rotating gas 
in the disk onto the remnant.  The idealized  model we adopt 
strictly applies only when the recoil speed of the central remnant is low, in 
which case the remant remains near the center of the hollowed disk as it 
accretes. For nonspinning black holes, the recoil speed is identically 
zero for equal-mass companions and small for companions of very unequal mass,
falling roughly as $q^2$, where $q<1$ is the mass ratio. It reaches
a maximum of about $175$ km/s for $q \approx 0.36$ (see \cite{BauS10} for
a review and references). For spinning
black holes of comparable mass the recoil can be much larger, but it is 
very sensitive to the spin magnitudes and orientations~\cite{RECOIL} 
and again decreases as $q^2$. Recoil speeds 
are estimated to be low ($< 200$ km/s) in most galactic mergers~\cite{BodRM07}.
Recent simulations of 
BHBHs orbiting inside circumnuclear disks in merged galaxy remnants give
a median recoil speed that is quite low, $< 70$ km/s~\cite{DotVPCRH09}. 
Determining the distribution of pre-merger BHBH 
masses, mass ratios, spins and spin orientations remains an area of 
active investigation.  We anticipate that in generic cases, both 
hydrodynamical (e.g shock) and secular (viscous) dissipation of energy 
are generated in the disk following decoupling and contribute to the radiation 
afterglow. We also anticipate that turbulent magnetic fields will provide 
the main source of (effective) viscosity in a realistic accretion disk.
Future simulations will therefore need to incorporate magnetic fields to 
treat all phases of of the afterglow properly. 

We adopt geometrized units and set $G=1=c$ below.

\section{Basic Model}

We focus on the epoch following the decoupling of the binary BHBH from the
circumbinary accretion disk. Our calculation begins when the inner radius of
the disk reaches the decoupling radius $r_1$ and we follow its evolution as it 
fills in the hollow  around the merged remnant, $M$. We adopt Newtonian
physics throughout.  We assume that just prior to decoupling 
the balance between tidal torques from the binary 
and viscous stresses in the disk results in a near-equilibrium disk which has 
the profile of a stationary thin disk with an ``effective ISCO'' at $r_1$. We 
take the outer boundary of the disk to be infinite. This constitutes our
initial data.

As the gas loses angular momentum and diffuses inward from $r_1$ following 
decoupling, we assume that it continues to inspiral in Keplerian 
circular orbits about the central black hole in the thin-disk approximation. 
We adopt axisymmetry about the rotation axis of the disk and neglect
the self-gravity of gas in the disk.
We take the inner boundary of the inspiraling disk to
be the remnant black hole ISCO, $r_{\rm isco}$, where we assume 
that viscous torques vanish. 

In our model we assume that the energy dissipated by turbulent viscosity
in the disk is all emitted
as thermal black body radiation at the local effective
temperature. Thus the net flux is a time-dependent superposition of
local black body spectra from concentric annuli in the evolving disk, both
inside and outside $r_1$. The {\it magnitude} of the 
integrated (total) radiated flux 
does not depend on the black body assumption and is independent of the 
detailed microphysics governing the (vertical) structure of the disk, including 
its local scale height, viscosity, pressure, temperature, 
and opacity.  In an illustrative example we set the turbulent
viscosity $\nu$ to be constant,
noting that in some $\alpha$-disk models the viscosity is
found to be a weak function of radius~\cite{MilP05}.
The magnitude of the viscosity, which does depend on the adopted
microphysics, establishes the physical timescale for the 
inward diffusion of gas in the disk, but it can be scaled out of the
numerical evolution of the disk and the emission.  The disk microphysics is 
also important
for determining standard ``modifications'' to the black body spectrum
arising from, e.g., the dominance of electron scattering over true absorption 
in the inner region of the disk, as well as from comptonization or
synchrotron radiation~\cite{ADAF}. These effects tend to 
increase the color temperature above the effective temperature of the 
radiation and thereby harden the emitted spectrum.  
We postpone any consideration of these refinements to a future analysis.

\section{Basic Equations}
\subsection{Disk Structure}

The evolution equation for the surface density $\Sigma(t,r)$ in an axisymmetric,
geometrically-thin disk may be derived by combining the equation of mass 
conservation (continuity equation) and the equation of angular momentum 
conservation (the $\phi$-component of the Navier-Stokes equation). 
The result is~\cite{Pri81}
\begin{equation} \label{sigma}
\frac{\partial \Sigma}{\partial t} = \frac{3}{r} \frac{\partial}{\partial r}
\left( 
r^{1/2} \frac{\partial}{\partial r} \left( \nu \Sigma r^{1/2} \right)
\right)\ , 
\end{equation}
where $r$ is the cylindrical radius measured from the disk axis (or black hole
remnant), $t$ is the time and $\nu$ is the
viscosity, which can be quite general at this point (i.e., $\nu = \nu (t,r)$). 
Equation~(\ref{sigma})
assumes that the angular velocity of the disk is 
Keplerian, $\Omega = (M/r^3)^{1/2}$. Taking the accretion rate at infinity to 
be independent of time, we seek to solve
the equation subject to the following boundary conditions:
\begin{equation}
\label{bc}
\mbox{b.c.'s}: \ \ \ \nu \Sigma = \left \{ \begin{array}{ll}
(\nu \Sigma)_{\infty}={\rm constant},&  
r \rightarrow \infty \\ 
 0,& r = r_{\rm isco}
\end{array} \right. .
\end{equation}

For initial conditions, we shall take the system immediately after black 
hole merger at the center of the disk hollow. At merger the 
disk has not changed appreciably from its structure
at decoupling, given that the viscous timescale is much longer than the
BHBH inspiral timescale after decoupling.  Prior to decoupling 
the balance between 
tidal torques and viscous stresses produces a near-equilibrium disk with 
an ``effective ISCO'' at $r_1 = r_{\rm edge} \sim 1.5a - 2a$, where $a$ is the 
semimajor axis of the binary. Consequently,
we take the initial disk to be an equilibrium solution to 
equation~(\ref{sigma}) (i.e. a solution to $\partial \Sigma / \partial t = 0$) 
satisfying the boundary conditions
\begin{equation}
\label{initbc}
t=0: \ \ \ \nu \Sigma = \left \{ \begin{array}{ll}
(\nu \Sigma)_{\infty},&  r \rightarrow \infty \\ 
0,& r = r_1
\end{array} \right. .
\end{equation}
The initial profile of the disk is therefore a ``donut'' described by
\begin{equation} 
\label{initeq}
t=0: \ \ \ \nu \Sigma = \left \{ \begin{array}{ll}
(\nu \Sigma)_{\rm \infty} 
\left( 1 - \left(r_1/r \right)^{1/2} \right), & r \geq r_1 \\ 
            0,&  r < r_1  
\end{array} \right. .
\end{equation}

In fact, the profile of the final equilibrium disk, formed once
the ``donut'' hole fills with gas down to $r_{\rm isco}$ and settles down, 
has a similar, familiar form, 
\begin{equation}
\label{finaleq}
t=\infty: \ \ \ \nu \Sigma =  \left \{ \begin{array}{ll}
(\nu \Sigma)_{\rm \infty}
\left( 1 - \left( r_{\rm isco}/r \right)^{1/2} \right), 
 & r \geq r_{\rm isco} \\
           0,&  r < r_{\rm isco}
\end{array} \right. .
\end{equation}
The goal is to solve equation~(\ref{sigma}) for the transient evolution 
during the interval $0 < t < \infty$. 

The accretion rate $\dot{M}(t,r) = 2 \pi r \Sigma v_r$, where $v_r$ is the
inward radial velocity of the gas, may be calculated from
\begin{eqnarray} 
\label{mdot}
\dot{M}(t,r) &=& 
 - \left[ \frac{\partial (r^2 \Omega)}{\partial r} \right]^{-1}
            \frac{\partial G}{\partial r}, \cr
       &=&     6 \pi r^{1/2} \frac{\partial (r^{1/2} \nu \Sigma)}{\partial r}.
\end{eqnarray}
In equation~(\ref{mdot}) the quantity $G(t,r)$ is the viscous 
torque~\cite{Pri81},
\begin{equation}
G(t,r) = 2 \pi r^3 \nu \Sigma \frac{\partial \Omega}{\partial r}\ ,
\end{equation}
evaluated for Keplerian motion. Using equation~(\ref{bc}) to evaluate
the accretion rate as $r \rightarrow \infty$ yields
\begin{equation}
\label{mdotinf}
\dot{M}_{\infty} \equiv \dot{M}(t, \infty) = 3 \pi (\nu \Sigma)_{\infty} = {\rm constant}\ .
\end{equation}
Equation~(\ref{mdotinf}) relates the asymptotic accretion rate to the
asymptotic disk structure, both of which remain constant during the 
transient. This scenario is consistent with the notion that conditions
far away from the central binary control the rate at which gas is
fed into the disk.

To solve equation~(\ref{sigma}), which in general is nonlinear, it is 
convenient to introduce the following nondimensional variables:
\begin{eqnarray}
\label{nondim}
s &=& (r/r_1)^{1/2}, \ \ \bar{\Sigma} = \Sigma/\Sigma_{\infty}, \ \
y= s \bar{\Sigma}, \cr
\bar{\nu}&=&\nu/\nu_{\infty}, \ \ \tau=t/t_{\rm vis}\ , 
\end{eqnarray}
where
\begin{equation}
\label{tvis}
t_{\rm vis} = \frac{4}{3} \frac{r_1^2}{\nu_{\infty}}
\end{equation}
is a viscous timescale. In terms of these variables, 
equations~(\ref{sigma}) becomes 
\begin{equation}
\label{sigma2}
\frac{\partial y}{\partial \tau} = \frac{1}{s^2} \frac{\partial^2 \left( 
y \bar{\nu} \right)} {\partial s^2}\ ,
\end{equation}
and must be integrated subject to boundary conditions
\begin{equation}
\label{bc2}
\mbox{b.c.'s}: \ \ \ y =\left \{ \begin{array}{ll}
 s, & s \rightarrow \infty\, \\
 0, & s = \left( r_{\rm isco}/r_1 \right)^{1/2}
\end{array} \right. 
\end{equation}
and initial conditions
\begin{equation}
\label{init2}
t=0: \ \ \ \ y = \left \{ \begin{array}{ll} 
\left( s-1 \right)/\bar{\nu},  & s \geq 1 \\
         0, & s < 1
\end{array} \right. .
\end{equation}

\subsection{Afterglow Radiation}

The rate of viscous dissipation per unit surface area of the disk
is given by~\cite{Pri81}
\begin{equation}
\label{diss}
D_{\rm vis}(t,r)=\frac{9}{8}\nu \Sigma \frac{M}{r^3}.
\end{equation}
Evolution equation~(\ref{sigma}) or (\ref{sigma2}) must be solved 
to determine the transient dissipation rate~(\ref{diss}) as the disk fills up 
the hollow and then causes the region outside the original hollow to 
adjust in response.
Assuming that this energy is all radiated locally, and approximating the 
emission to be a thermal blackbody radiation,
the local disk surface temperature $T_s(t,r)$ may be equated to the 
effective temperature, which is determined by the 
dissipation rate according to
\begin{equation}
\label{Tsurf}
\sigma T_s^4(t,r)=D(t,r)=\frac{9}{8}\nu \Sigma \frac{M}{r^3}\ , \ \ \ 
0 \leq t \leq \infty\ ,
\end{equation}
where $\sigma$ is the Stefan-Boltzmann constant.
For the initial and final equilibrium disks described by 
equations~(\ref{initeq}) and ~(\ref{finaleq}) the result can be found
immediately:
\begin{equation}
\label{Tseq}
\sigma T_s^4(t_{\rm eq},r) = \left \{ \begin{array}{ll}
\left(3 M \dot{M}_{\infty}/ 8 \pi r^3 \right)
\left( 1 - \left( r_{\rm eq}/r \right)^{1/2} \right), & r \geq r_{\rm eq} \\
0, &  r < r_{\rm eq} 
\end{array} \right. ,
\end{equation}
where 
\begin{equation}
\label{req}
t_{\rm eq}= \left \{ \begin{array}{ll}
0,  & r_{\rm eq}=r_1, \\
\infty, & r_{\rm eq}=r_{\rm isco}
\end{array} \right. .
\end{equation}
For the transient, the temperature must be determined numerically by
integrating equation~(\ref{sigma2}) and substituting the result into
equation~(\ref{Tsurf}).

Given the surface temperature, the transient specific flux $F_{\nu}(t)$ 
measured by an observer at 
distance $d$ whose line of sight makes an angle $i$ to the normal to the disk 
plane is determined by integrating over the entire disk surface, 
\begin{equation}
\label{flux1}
F_{\nu}(t) = \frac{2 \pi ~{\rm cos}~i} {d^2} 
\int_{r_{\rm isco}}^{\infty} B_{\nu}(T_s(t^{\prime},r)) r dr, 
\end{equation}
where  $B_{\nu}(T_s(t^{\prime},r))$ is the Planck function, $t^{\prime} 
= t-d$ is retarded time and $\nu$ is the photon frequency~\cite{FREQ}.
Equation~(\ref{flux1})
is best evaluated in terms of a nondimensional function 
$f^*(t^{\prime},x)$ of nondimensional frequency $x$ according to
\begin{equation}
\label{flux}
F_{\nu}(t) =  \frac{2 \pi ~{\rm cos}~i}{d^2} \frac{15}{\pi^5} \frac{\sigma T_*^4}
{\nu_*} r_{\rm isco}^2 f^*(t^{\prime},x),
\end{equation}
where
\begin{equation}
\label{f*}
f^*(t^{\prime},x) \equiv \int_1^{\infty} du ~u \frac{x^3}
{{\rm exp} \left(x T_*/T_s \right) -1},
\end{equation}
and where we have introduced the parameters
\begin{eqnarray}
\label{T*}
\sigma T_*^4 &\equiv& 3 M \dot{M}_{\infty}/8 \pi r_{\rm isco}^3, 
\ \ \ h \nu_* \equiv kT_*, \cr
x &\equiv& h \nu /k T_* = \nu/\nu_*, \ \ \ u \equiv r/r_{\rm isco}.
\end{eqnarray}
The quantity $T_*$ provides an estimate of the characteristic 
temperature in the main radiating region of the final equilibrium 
disk, which resides near $r_{\rm isco}$, and $h \nu_*$ is the 
characteristic frequency of the emitted thermal radiation from this region.  
The specific luminosity $L_{\nu} (t)$, summing over both surfaces of the disk,
is related to $F_{\nu}(t)$ according to
\begin{equation}
L_{\nu} (t) = \frac{2 \pi d^2}{{\rm cos}~i} F_{\nu}(t).
\end{equation}
The total luminosity $L(t)$ integrated over all frequencies is then given by
\begin{equation}
\label{lum1}
L(t) = \int_0^{\infty} d \nu L_{\nu}(t) = \frac{60}{\pi^3} \sigma T_*^4 
r_{\rm isco}^2 \int_0^{\infty} f^*(t^{\prime},x) dx
\end{equation}

The the initial and final equilibrium disks provide analytic temperature 
profiles (i.e. equation~\ref{Tseq}), from which the fluxes can be determined 
without knowing the transient evolution. At low frequencies 
$x \ll (r_{\rm isco}/r_{\rm eq})^{3/4}$, 
where $r_{\rm eq}$ is defined in equation~(\ref{req}),  we have
\begin{equation}
\label{lowfreq}
f^*(t_{\rm eq},x) \approx \frac{4}{3} \Gamma(8/3) \zeta(8/3) x^{1/3},
\end{equation}
where $\Gamma(8/3) = 1.50457$ and $\zeta(8/3) = 1.28419$. To derive this 
result we have used the approximation 
$T_*/T_s(t_{\rm eq,r}) \approx (r/r_{\rm isco})^{3/4}$ 
in the outer disk where the low-frequency radiation is generated, 
together with the identity
\begin{equation}
\Gamma(s) \zeta(s) = \int_0^\infty \frac{x^{s-1}}{{\rm exp}(x)-1}dx.
\end{equation}
Equations~(\ref{flux}) and (\ref{lowfreq}) combine to give
the well-known low-frequency component of the thin-disk thermal spectrum, 
$F_{\nu} \sim \nu^{1/3}$. 
The total integrated luminosity and flux from the initial and final disk
also can be obtained analytically. Using the identity
\begin{equation}
\int_0^{\infty} f^*(t^{\prime}_{\rm eq},x) dx = \frac{\pi^4}{45} 
\frac{r_{\rm isco}}{r_{\rm eq}}
\end{equation}
in equation~(\ref{lum1}) gives
\begin{equation}
L(t_{\rm eq}) = \frac{4}{3}\pi \sigma T_*^4 r_{\rm isco}^2 
\frac{r_{\rm isco}}{r_{\rm eq}}
\end{equation}
or, equivalently,
\begin{equation}
\label{Ltot}
L(t_{\rm eq}) = \left \{ \begin{array}{ll}
M \dot{M}_{\infty}/2 r_1~, & t^{\prime}_{\rm eq}=0, \\
M \dot{M}_{\infty}/2 r_{\rm isco}~, & t^{\prime}_{\rm eq}=\infty 
\end{array} \right. .
\end{equation}
Equation~(\ref{Ltot}) recovers the standard result for the total luminosity
of a thin accretion disk with an inner boundary at $r_{\rm eq}$. This result
is customarily obtained directly from the integral
\begin{equation}
\label{Ltot2}
L(t) = 2 \int_{r_{\rm eq}}^{\infty} \sigma T_s^4 ~2 \pi r dr
=  2 \int_{r_{\rm eq}}^{\infty}\frac{9}{8}(\nu \Sigma) \frac{M}{r^3} 
~2 \pi r dr,
\end{equation}
substituting equations~(\ref{initeq}) and (\ref{finaleq}) for the two
equilibrium profiles.
Its significance for afterglow radiation is that, at least in our 
simple model, the total electromagnetic flux will increase 
from its value immediately after
BHBH merger (as signaled by peak gravitational wave emission) to a value 
that is larger by a factor of $r_1/r_{\rm isco}$ as the disk settles into
final equilibrium. The timescale over which the increase will occur will 
be the viscous timescale near the decoupling radius in the disk, $r_1$.
In the next section we calculate the time-varying evolution 
and electromagnetic emission during this transition.

\section{The Transient Afterglow: An Illustrative Example}

\subsection{Nondimensional Description}
 
Equation~(\ref{sigma2}) is a parabolic, nonlinear, diffusion equation 
with a variable
diffusion coeffient (e.g., the coefficient decreases as $s^{-2}$ for
the linear case with constant $\bar{\nu}$). In general,
such a nonlinear equation must be integrated
numerically, which, in anticipation of future applications, 
motivates our approach here. However, there are standard
analytic (e.g. Green function) techniques that can be implemented for the
linear case~\cite{Pri81}.  We adopt a Crank-Nicholson 
finite-difference scheme, which is fully implicit and second order 
in both space and time. Such an implicit scheme is well-suited to 
accommodate the large spatial extent of the disk that must be 
covered to implement the boundary conditions. This coverage is facilitated by 
introducing a logarithmic grid in $s$ from $s=(r_{\rm isco}/r_1)^{1/2} \ll 1$ 
to $ s \gg 1$~\cite{BC}.
Since a Crank-Nicholson scheme is unconditionally stable for numerical 
timesteps of arbitrary size, the step size is restricted only by
accuracy considerations and not grid spacing and can be much larger than that 
required by the usual Courant condition for explicit schemes.

To illustrate the transient behavior with a simple model,  
we take the viscosity in the disk to be a constant,
in which case $\bar{\nu} = 1$. Equation~(\ref{sigma2}) is then uniquely 
determined by setting $r_{\rm isco}$ and $r_1$. We set $r_{\rm isco} = 6M$,
the value appropriate for a Schwarzschild black hole remnant~\cite{ISCO}, 
and $r_1 = 100M$, a choice consistent with
some recent estimates of disk properties at 
binary BHBH-disk decoupling~\cite{DECOUP}. 

\begin{figure}
\includegraphics[width=9cm]{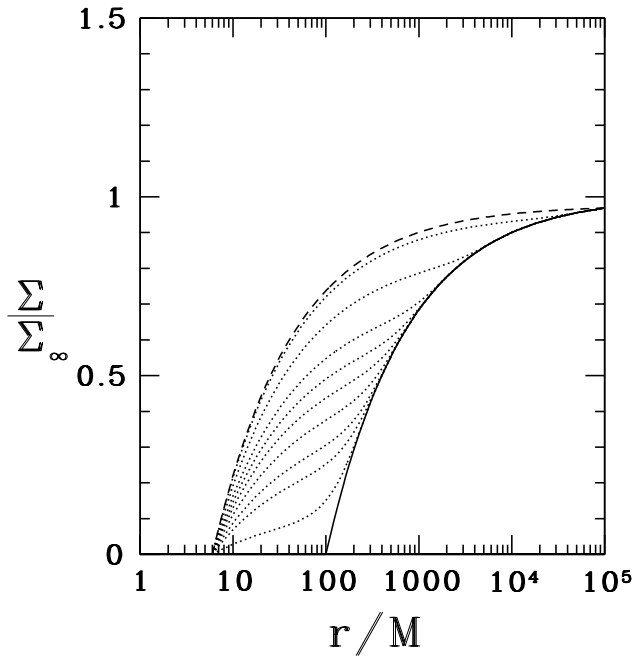}
\caption{Profiles of the disk surface density $\Sigma$ at select times 
$\tau$ during accretion following the binary BHBH merger. The 
disk decoupling radius is at $r_1 = 100 M$
and the ISCO is at $r_{\rm isco} = 6M$, where $M$ is the mass of the black 
hole remnat.  The initial (final) equilibrium disk is
indicated by the heavy solid (dashed) curve. Profiles at 
intermediate times are shown by dotted curves at $\tau = 0.0694, 0.208, 0.347,
0.694, 1.39, 2.78, 6.94, 69.4$ and $6940$.}
\label{fig:sigma}
\end{figure}

\begin{figure}
\includegraphics[width=9cm]{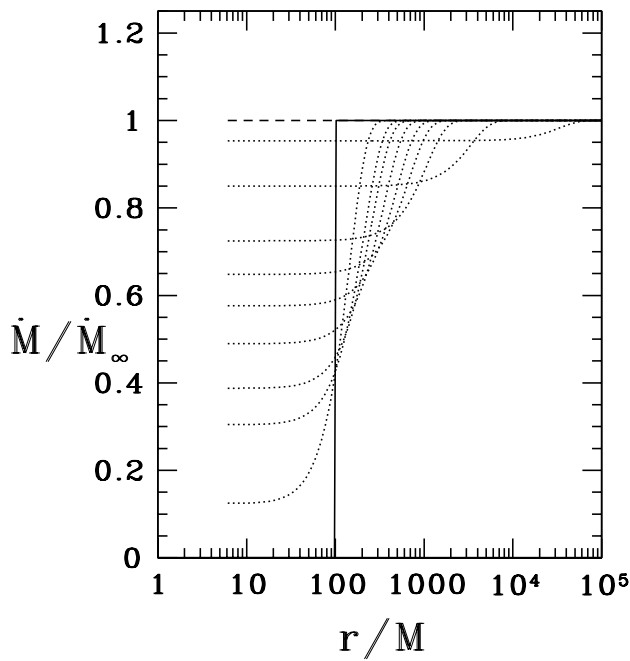}
\caption{Profiles of the local disk accretion rate $\dot{M}$ at select times
$\tau$ following the binary BHBH merger (see equation~\ref{mdot}). The curves are
labeled as in Fig.~\ref{fig:sigma}.}
\label{fig:mdot}
\end{figure}

The growth of the surface density $\Sigma(t,r)$ inside and outside 
the hollow following decoupling and merger is shown in 
Fig.~\ref{fig:sigma}.  The associated growth in the local accretion 
rate $\dot{M}(t,r)$ (see equation~\ref{mdot}) is 
plotted in Fig.~\ref{fig:mdot}.  The secular transition
from the initial to final equilibrium disk is evident as gas
migrates inward toward the black hole remnant.

\begin{figure}
\includegraphics[width=9cm]{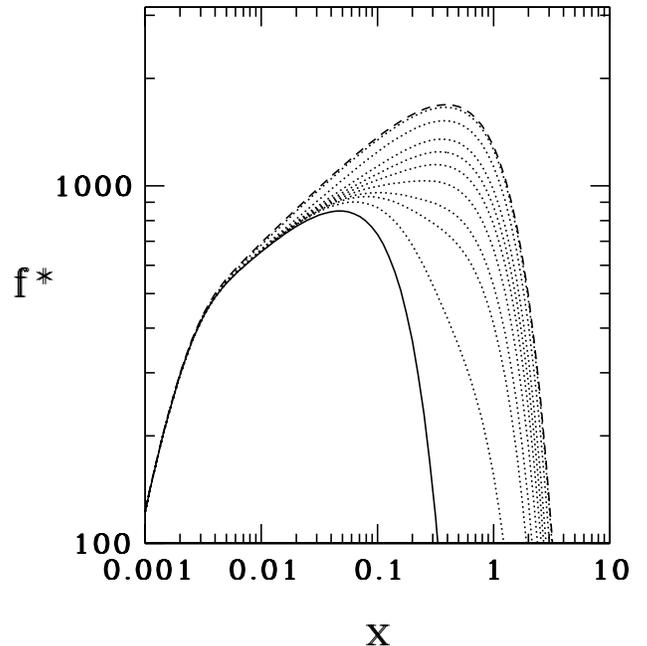}
\caption{The frequency distribution of the emitted electromagnetic thermal
radiation at select times $\tau$ following the binary BHBH merger. The 
nondimensional distribution function $f^*$ and photon frequency $x$ 
are defined in equations~(\ref{flux})--(\ref{T*}). 
The curves are labeled as in Fig.~\ref{fig:sigma}.}
\label{fig:flux}
\end{figure}

   The transient thermal spectrum (equations~\ref{flux} and \ref{f*}) 
is plotted in Fig.~\ref{fig:flux}. The spectrum 
hardens and the total flux and luminosity increase in magnitude as the
hollow fills with (radiating) matter.
In analogy to equation~(\ref{T*}), define the quantities
\begin{equation} \label{T1}
\sigma T_1^4 \equiv 3 M \dot{M}_{\infty}/8 \pi r_1^3,
\ \ \ h \nu_1 \equiv kT_1 
\end{equation}
which characterize conditions near the inner edge of the initial disk.
The peak frequency 
evolves from near $\nu_1$ to $\nu_*$ over the course of the 
transient, where
$\nu_*/\nu_1 = (r_1/r_{\rm isco})^{3/4} \approx 8.2$ in our example.
Over the same time frame, the total luminosity increases according to
equation~(\ref{Ltot}), whereby $L(\infty)/L(0) = r_1/r_{\rm isco} \approx 16.7$
in this example~\cite{TIME}.

\begin{figure}
\includegraphics[width=9cm]{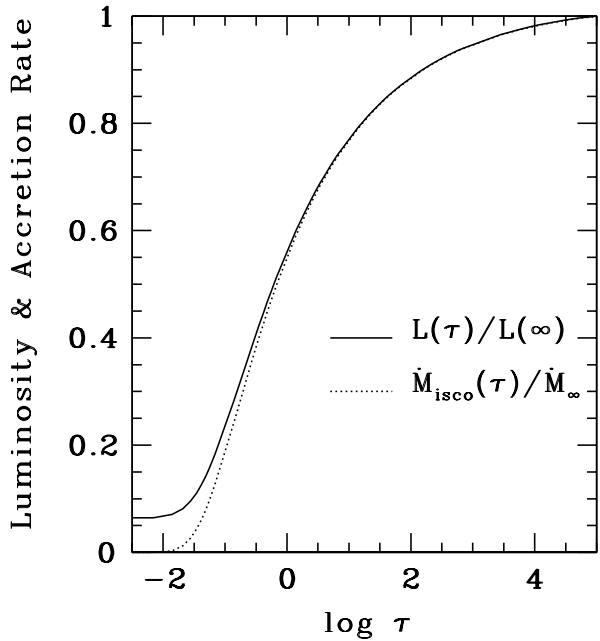}
\caption{The increase in luminosity $L$ and ISCO accretion rate 
$\dot{M}_{\rm isco}$ with time $\tau$ during accretion following the
binary BHBH merger. Note that the time coordinate for $\dot{M}_{\rm isco}$ 
represents retarded time.}
\label{fig:lum}
\end{figure}

The increase in total luminosity is plotted as a function
of time in Fig.~\ref{fig:lum}. The characteristic 
timescale over which the transient lasts is $\delta \tau \sim 10 - 100 $,
or, according to equations~(\ref{nondim}) and (\ref{tvis}), 
$\delta t \sim (10 - 100)  ~t_{\rm vis}$.  We note that the 
timescale $t_{\rm vis}$ defined by equation~(\ref{nondim}) is related to the
viscous timescale at the inner edge of initial disk, $r_1$, where the 
initial surface density, $\Sigma(0,r_1)$, is identically zero 
(equation~\ref{initeq}). The gas that establishes the final equilibrium 
flow at the ISCO originates from radii $r >> r_1$, 
where the the density $\Sigma(0,r)$ is nonzero and closer to its final 
equilibrium value. Since the viscous timescale on which
this gas drifts inward varies like $r^2$, the equilibration timescale is 
therefore much longer than $t_{\rm vis}$.

The luminosity increase with time is directly correlated 
with the increase in the 
accretion rate at the ISCO, as is evident from 
Fig.~\ref{fig:lum}. That the two curves plotted in the figure
line up so closely once the transient gets underway ($\tau \gtrsim 0.1)$ 
can be explained as follows: during this epoch, 
the surface density in the main radiating, inner region is found 
numerically to satisfy the approximate relation
\begin{equation}
\label{transient}
\nu \Sigma(t,r) \approx 
C(t) \left( 1 - (r_{\rm isco}/r)^{1/2} \right),
\end{equation}
where the factor $C(t)$ climbs from near zero initially to its 
late, asymptotic value $C(\infty)=(\nu \Sigma)_{\infty}$ 
(see equation~\ref{finaleq}). Using equation~(\ref{transient}) in
equation~(\ref{Ltot2}) yields
\begin{equation}
\label{L(t)}
L(t)/L(\infty) \approx C(t^{\prime})/(\nu \Sigma)_\infty.
\end{equation}
Using equation~(\ref{transient}) in equation~(\ref{mdot}) to calculate 
$\dot{M}_{\rm isco}(t)$, and 
inserting equation~(\ref{mdotinf}), then gives
\begin{equation}
\dot{M}_{\rm isco}(t^{\prime})/\dot{M}_{\infty} \approx L(t)/L(\infty),
\end{equation}
which explains the numerical result revealed in Fig.~\ref{fig:lum}.
Consistent with the fit (\ref{transient}) for the surface density 
is the implication that the accretion rate in innermost region near
the ISCO must be nearly independent of $r$ during the transient, 
$\dot{M}(t,r) \approx 3 \pi C(t)$. This result is  
verified numerically in Fig.~\ref{fig:mdot}.

\subsection{Physical Units and Scaling}

To be astrophysically useful and applicable to general cases
we need to restore physical units to our nondimensional numerical 
solution and provide appropriate scaling with the 
independent physical parameters.  
Toward this end we define several auxiliary quantities, including 
the black hole mass parameter
$M_6 = M/10^6 M_{\odot}$, the distance parameter $d_{10} = d/10$ Mpc, 
and the accretion rate in Eddington units, 
$\dot{m}_E = \dot{M}/\dot{M}_E$, where 
$\dot{M}_E = 12 L_E/c^2 = 1.68 \times 10^{24} M_6$ gm/s is the 
Eddington accretion rate for the final equilibrium
disk~\cite{EDD} and $L_E = 1.26 \times 10^{44} M_6$ erg/s is the 
Eddington luminosity. In terms of these parameters, we have
\begin{eqnarray}
\label{scale1}
T_* &=& 0.906 \times 10^6 ~\dot{m}_E^{1/4} M_6^{-1/4} \  {\rm K}, \cr
\nu_* &=& 1.89 \times 10^{16} ~\dot{m}_E^{1/4} M_6^{-1/4} \  {\rm Hz}, \cr
      &=& 78.1 ~\dot{m}_E^{1/4} M_6^{-1/4} \ {\rm eV/h}, \cr
\frac{\sigma T_*^4 r_{\rm isco}^2}{\nu_* d^2} &=& 
1.67 \times 10^{-24} ~\dot{m}_E^{3/4} M_6^{5/4}d_{10}^{-2} \ 
{\rm \frac{erg}{cm^2~s~Hz}}, \cr 
L(\infty) &=& 1.26 \times 10^{44} ~\dot{m}_E \ {\rm erg/s}.
\end{eqnarray}
The above quantities set the scale for the characteristic 
innermost disk temperature, peak thermal photon frequency, observed 
thermal flux and total thermal luminosity for the final 
equilibrium disk with $r_{\rm isco}=6M$.
The fourth line provides the scale factor that appears in 
equation~(\ref{flux}) for the specific flux. Using this factor together 
with Fig.~\ref{fig:flux} yields the the transient 
spectrum in physical units.

The corresponding parameters for the initial disk at
decoupling, with $r_1 = 100M$, are given by
\begin{eqnarray}
\label{scale2}
T_1 &=& 0.653 \times 10^5 ~\dot{m}_E^{1/4} M_6^{-1/4} \  {\rm K}, \cr
\nu_1 &=& 1.36 \times 10^{15} ~\dot{m}_E^{1/4} M_6^{-1/4} \  {\rm Hz}, \cr
      &=& 5.63 ~\dot{m}_E^{1/4} M_6^{-1/4} \ {\rm eV/h}, \cr
\frac{\sigma T_1^4 r_1^2}{\nu_1 d^2} &=&
0.695 \times 10^{-24} ~\dot{m}_E^{3/4} M_6^{5/4}d_{10}^{-2} \
{\rm \frac{erg}{cm^2~s~Hz}}, \cr
L(0) &=& 0.754 \times 10^{43} ~\dot{m}_E \ {\rm erg/s}.
\end{eqnarray}
The fact that the thermal emission hardens and increases in magnitude
during the transient is again evident by comparing equations~(\ref{scale1}) and
(\ref{scale2}). For black holes of mass $M_6 \approx 1$ accreting at
$\dot{m}_E \approx 0.1$ the peak emission at the start of the transient
lies in the $UV$ band and then hardens to $EUV$ and soft $X$-ray 
radiation at late times~\cite{COLOR}.

None of the physical disk structure or emission frequencies and magnitudes 
calculated  above depend on any details of the disk microphysics 
in our idealized model.  However, we do need to probe the microphysics 
in order to assign a physical {\it timescale} to the transient. The
dimensional time parameter $\tau$ can be translated into physical time
only when $t_{\rm vis}$ appearing in equation~(\ref{tvis}) has been evaluated,
and for that we need to know the viscosity. 
Our numerical implementation
assumed a constant viscosity in the disk. To get a realistic estimate of
the appropriate value to use we can  
adopt the standard Shakura and Sunyaev~\cite{ShaS73} $\alpha$-disk model 
and employ the value it gives for the turbulent viscosity 
at $r_1$. In this model, the
radius $r_1$ will reside in the ``inner region'', 
where the pressure is radiation-dominated and the opacity is 
electron scattering-dominated, provided it lies within a critical radius 
$r_{\rm in}$, where~\cite{NovT73,ShaT83}
\begin{equation}
r_{\rm in}/M \approx 
2.56 \times 10^3 ~{\alpha}^{2/21} \dot{m}_E^{16/21} M_6^{2/21}.
\end{equation}
For those cases in which $r_1 < r_{\rm in}$~\cite{MilP05} 
the viscous timescale at $r_1/M= 100$ is  
\begin{equation}
t_{\rm vis} \approx 2.54 \times 10^5 ~{\alpha}^{-1} \dot{m}_E^{-2} M_6 \
{\rm sec}.
\end{equation}
Hence for  the case $M_6=1, \dot{m}_E = 0.1$ and $\alpha = 0.1$ we obtain
$t_{\rm vis} = 8.1$ yr. 
For other cases, $r_1$ may lie in the gas-pressure dominated
region and the estimate for $t_{\rm vis}$ will be quite different~\cite{SchK08}.

In reality, secular gas inflow in the hollow following BHBH merger 
is likely controlled by magnetic fields that are driven turbulent by
MRI and other instabilities~\cite{MRI}. The magnetic field thereby acts like
an effective turbulent viscosity, but such a viscosity 
might exhibit very different scaling behavior 
from the standard $\alpha-$disk model. This issue, which is
the subject of considerable interest and current research in the case 
of stationary disk accretion onto black holes, is clearly relevant in the
case of transient, post-merger BHBH accretion, but it is beyond the 
scope of this paper and we shall not pursue it further.

\subsection{Super-Eddington Accretion?}

The evolution calculated above assumes that the rate at which  
gas enters the disk is determined by conditions at large distance 
from the remnant black hole and remains constant in time during 
the transient. In cases for which 
the ambient gas densities are sufficiently low the merging BHBH may be
`gas-starved' and the accretion rate will be safely sub-Eddington during the
entire evolution.  However, an interesting scenario arises if there is 
sufficient gas in the neighborhood of the merging (`gas-rich') BHBH 
for the luminosity to exceed the critical Eddington limit $L_E$.
Suppose, for example, that already by the time of merger the 
accretion rate $\dot{m}(0)$ at $r_1$ in the initial hollowed-out disk 
is sufficiently high to drive the luminosity $L(0)$ to the critical 
Eddington value $L_E$. Then  accretion at this same rate would cause the
luminosity to exceed $L_E$  during the subsequent transient phase, as the 
total disk luminosity steadily increases (see Fig.~\ref{fig:lum}). 
According to equation~(\ref{Ltot}) this initial accretion rate will be
larger than the critical Eddington rate in the final equilibrium disk by 
a substantial factor: 
$\dot{m}(\infty) = \dot{m}(0) = (r_1/r_{\rm isco})\dot{m}_E > \dot{m}_E $.  
Ignore for the moment the possibility that the geometrically 
thin-disk approximation could break down as $L$ approaches 
$L_E$~\cite{FraKR02}.  Since the viscous timescale
increases rapidly with $r$ (see, e.g., equation~\ref{tvis})
the outer disk has insufficient time to adjust quasistatically and
reduce the inflow rate to maintain $L(t)$ at $L_E$ as 
gas fills the hollow.  What happens instead is unclear and 
may be complicated and highly dynamical.  One possibility is 
that the disk puffs up, the accretion becomes almost spherical as the 
gas fills the hollow, and the local radiation pressure blows excess matter away 
to maintain $L \approx L_E$.  Alternatively,  density inhomogeneities 
resulting from the photon-bubble instability~\cite{Aro92,Gam98}
might keep the disk geometrically thin and allow radiation to escape through
porous regions of very low density. In the latter case the escaping 
luminosity could exceed $L_E$ by a considerable factor ($10-100$),
accompanied by radiation-driven mass loss~\cite{Beg02}.  
A full understanding of this super-Eddington scenario may
require global, general relativistic, radiation-MHD simulations in 
$3+1$ dimensions. Such a treatment may well be worthwhile, since this 
transition from quiescent sub-Eddington accretion to highly dynamical 
super-Eddington flow may provide another observable signature of 
binary BHBH mergers in `gas-rich' regions.

{\it Acknowledgments}: It is a pleasure to thank 
C. Gammie and Y.T. Liu for useful discussions and D. Kotan for 
technical assistance.  This paper was supported in part by NSF
Grant PHY06-50377 and NASA Grant NNX07AG96G to the University of Illinois 
at Urbana-Champaign.

\bibliography{paper3.bib}

\begin{thebibliography}{53}
\bibitem[{ADA()}]{ADAF}
\bibinfo{note}{{Details involving the disk microphysics are also required to
  estimate the fraction of the dissipated energy that may be advected rather
  than radiated; we neglect advection here.}}
\bibitem[{FRE()}]{FREQ}
\bibinfo{note}{{From the context it should not be difficult to distinguish the
  photon frequency $\nu$ from the viscosity $\nu$.}}
\bibitem[{BC()}]{BC}
\bibinfo{note}{{While the disk is presumed to be infinite, in practice we set
  its outer boundary at $r = 10^5 M$ in our numerical integrations; the results
  are insensitive to the specific choice of outer radius for values this far
  out in the disk.}}
\bibitem[{ISC()}]{ISCO}
\bibinfo{note}{{However, the remnant of two equal-mass, nonspinning BHBHs will
  be a rotating Kerr black hole with dimensionless spin parameter $J^2/M
  \approx 0.688$, for which $r_{\rm isco}/M \approx 3.45$ in Boyer-Lindquist
  coordinates; see, e.g.,~\cite{BauS10}.}}
\bibitem[{DEC()}]{DECOUP}
\bibinfo{note}{{See, e.g., \cite{MilP05}, who find that fiducial disk
  parameters give a decoupling binary separation $a \approx 100M$, scaling
  slowly with mass as $\left( M/10^6 M_{\odot} \right)^{0.08} \left( 4q/(1+q)^2
  \right)^{0.42}$. See also \cite{SchK08}, who find higher values of $a$.}}
\bibitem[{TIM()}]{TIME}
\bibinfo{note}{{Here our notation $L(0)$ and $L(\infty)$ refers to the initial
  and final equilibrium disks, respectively. }}.
\bibitem[{EDD()}]{EDD}
\bibinfo{note}{{See equation~(\ref{Ltot}), which is for a Newtonian disk and
  gives an accretion efficiency $\eta=1/12$ for $r_{\rm isco} = 6M$. }}.
\end{thebibliography}
\end{document}